\documentclass[a4paper,11pt]{article}
\usepackage[usenames]{color}

\usepackage{amsfonts,amssymb}
\usepackage{theorem}
\usepackage[dvips]{graphicx}
\newcommand{\su}[2]{\stackrel{#1}{#2}}

\begin{document}
\rightline{}
\vskip 3 truecm
\par
\normalsize
\centerline{\bf On the Renormalization of the Complex Scalar Free Field Theory
\footnote{\tt This work is supported in part by funds provided by the 
U.S. Department of Energy (D.O.E.) under cooperative research 
agreement \#DE FG02-05ER41360}}
\normalsize
\rm 
\large
\vskip 1.5 truecm
\centerline{
Ruggero ~Ferrari
\footnote{e-mail: {\tt ruggferr@mit.edu}}}
\small
\begin{center}
Center for Theoretical Physics\\
Laboratory for Nuclear Science
and Department of Physics\\
Massachusetts Institute of Technology\\
Cambridge, Massachusetts 02139\\
and\\
Dip. di Fisica, Universit\`a degli Studi di Milano\\
and INFN, Sez. di Milano, Italy\\
(MIT-CTP-4045, IFUM-940-FT, July  2009)
\end{center}
%
%
%

\vskip 1.5 truecm
\normalsize
\bf
\centerline{Abstract}

\rm
\begin{quotation}
Polar coordinates are used for the complex   scalar free field
in $D=4$ dimensions.
The resulting non renormalizable theory is healed by using a recently
proposed symmetric subtraction procedure. The existence of the
coordinates transformation is proved by construction.

\end{quotation}
\newpage
\section{Introduction}
\label{sec:intr}
In power counting renormalizable theories there is a universally
accepted rule, by which to every independent divergent one-particle-irreducible amplitude
(1PI)
one must associate a  parameter in the tree-level action.
This rule cannot be easily exported to any program of subtraction of
infinities in nonrenormalizable theories. In fact, if this rule is used, 
the theory looses (in general) any predictivity and moreover the perturbative approach is
unstable: for every new divergent 1PI amplitude emerging
in the perturbative expansion, the whole
series have to be updated from the beginning. 
Because of these two reasons we have proposed
a new approach to the subtraction procedure for non renormalizable theories
\cite{Ferrari:2005ii}, \cite{Bettinelli:2007zn}.
To our opinion the removal of the infinities have to be considered as a pure mathematical
problem that aims to give a meaning to undefined expressions. Obvious properties
have to be maintained as locality of the counterterms and physical unitarity.
\par
We have proposed a subtraction strategy where
the symmetry properties of the path integral measure 
and the dynamics are imposed through  Local Functional Equations (LFE's)
obeyed by the connected functional or by the vertex functional 
\cite{Ferrari:2005ii}. The action
is not an adequate quantity in this procedure. This strategy of subtraction
has been thoroughly analyzed \cite{analysis}
and successfully used for the nonlinear sigma model \cite{sigma},
for massive YM theory \cite{ym},  and for the Electroweak Model 
\cite{ew}.
\par
Persistent objections from
some experts in the field about this new strategy of ours has led me to consider a crystal
clear (hopefully
\footnote{The presence of second derivatives in the interaction vertices
(e.g. equations (\ref{conv.1}) and (\ref{conv.2}))
causes some problems: 1PI functions do not coincide with the Legendre transform
of the connected amplitudes. In fact the equation of motion might eventually
contract an internal line to a point. 
However this seems not to affect the subtraction procedure 
here presented. In particular one-particle-reducible graphs, promoted to
1PI by the equation of motion, enjoy the property of factorization
and therefore they do not require extra subtractions.}
) example: free field theory. The example turned out to be much
more interesting than I thought and therefore I decided to write it down.
The state of the art is difficult to tell, since this problem is as old
as quantum field theory and it is strictly connected to that of field-coordinate
transformations. Therefore I apologize for the missed references. 
\par
The paper is self contained; however the proofs are only sketched, being
present in previous works. I shall argue that standard polar coordinates 
cannot be used since the
perturbative expansion (in loops) is around a vacuum where 
Spontaneous Breakdown of the $U(1)$ Symmetry of the complex field  
occurs and a Goldstone boson appears.  
I modify the polar coordinates in order to meet the basic 
requirements of the equivalence theorem \cite{dyson}-\cite{Kamefuchi:1961sb}.
Then  I show how 
the local symmetry transformations, associated to the path integral measure, 
can be implemented by using an
infinite set of external sources, which eventually appear in the LFE's. 
LFE's are then used to prove the hierarchy structure of the vertex
functional and finally to establish the subtraction procedure.
It is amazing how the free field structure remains in such 
a complicated non renormalizable theory. Finally  the sturdy Sections 
(\ref{sec:sol}-\ref{sec:local}) are devoted to study the general 
structure of the counterterms,
by solving the LFE's at one-loop level, and to show why
the textbook renormalization cannot manage the polar coordinates
transformation. In Section \ref{sec:hlc} a solution of the LFE's
for the two-loop case is derived.
\par\noindent 
  I leave to the conclusions a detailed discussion of the results.


\section{Modified  Polar Coordinates  }
\label{sec:polar}
%

 I consider the action
\begin{eqnarray}
{\cal S}
= \Lambda^{D-4}\int d^D x [\partial_\mu \phi^*\partial^\mu  \phi
-m^2 \phi^*\phi ].
\label{intr.1}
\end{eqnarray}
If  one uses the following parameterization for polar coordinates
\begin{eqnarray}&&
\phi = e^{\chi}
\nonumber\\&&
\chi = \psi + i \theta
\label{exp.1}
\end{eqnarray}
where  $\theta,\psi\in [-\infty,\infty]$,
the action becomes
\begin{eqnarray}&&
S= \Lambda^{D-4}\int d^Dx e^{\chi}e^{\chi^*}\Bigl[\partial^\mu \chi^* \partial_\mu \chi  -
m^2  \Bigr]
\nonumber\\&&
=\Lambda^{D-4}\int d^Dx e^{2\psi}\Bigl[\partial^\mu \psi \partial_\mu \psi
+\partial^\mu \theta \partial_\mu \theta-
m^2 \Bigr].
\label{exp.2}
\end{eqnarray}
A series expansion of the exponential and a subsequent perturbative
approach would  take to a theory with a massless
field $\theta$. Since  I want massive scalar field
the parameterization in eq. (\ref{exp.1}) is not a good one.
Instead  I use
\begin{eqnarray}
e^\chi(x)=v^{-1}\phi(x)+1
\label{infinite.1}
\end{eqnarray}
(where $v$ is a mass scale) so that the power expansion starts with
\begin{eqnarray}
\chi(x)\sim v^{-1}\phi(x)
\label{infinite.2}
\end{eqnarray}
and therefore the equivalence theorem can be applied.
The action in the chosen variables ($v$ might be absorbed by a overall
rescaling of mass, coordinates and $\Lambda$)
\begin{eqnarray}
S= \Lambda^{D-4}\int d^Dx \Bigl[e^{\chi}e^{\chi^*}\partial^\mu \chi^* \partial_\mu \chi  -
m^2  (e^{\chi^*}-1)(e^{\chi}-1)\Bigr]
\label{infinite.3}
\end{eqnarray}
which  I can split into a free and interaction action
\begin{eqnarray}&&
=\Lambda^{D-4}\int d^Dx \Bigl[\partial^\mu \chi^* \partial_\mu \chi
- m^2 \chi^*\chi\Bigr] 
\nonumber\\&&  
+ \Lambda^{D-4}\int d^Dx \Bigl[\sum_{n=1}^\infty\frac{1}{n!}\partial^\mu \chi^* \partial_\mu \chi \,\,
(\chi^*+\chi)^n
\nonumber\\&&
-m^2\Bigl( \sum_{n=2}^\infty\frac{1}{n!}\chi^* \,\,\chi^{n}
+ \sum_{n=2}^\infty\frac{1}{n!}\chi\,\,\,\chi^{*\, n}
+ \Bigl|\sum_{n=2}^\infty\frac{1}{n!}\chi^{n}\Bigr|^2
\Bigr)
\Bigr] .
\label{infinite.4}
\end{eqnarray}
The perturbative expansion is in $\hbar$ (i.e the number of loops,
with some care on counting the $\hbar$ powers of the counterterms).  I will try to give a meaning  to the infinite number of divergent 1PI amplitudes
by using dimensional regularization
and eventually recover the free field theory in the variables
$\phi, \,\,\, \phi^*$ at $D=4$.
%
%
%
\section{On the Conventional Approach}
\label{sec:conv}
By proceeding in the conventional way,  one meets
a series of difficulties that have discouraged
people to discuss the problem of coordinate
transformations in quantum field theories. There has been
some important progress in the use of field redefinition
and its relation with the renormalization procedure 
\cite{Lam:1973qa}, \cite{Bergere:1975tr} and with the
algebraic structure of the theory \cite{Alfaro:1989xx}-
\cite{Ferrari:2002kz}. However, to my knowledge, no one
has directly faced the task of taming the difficulties
urging from  the arbitrariness of the counterterms in conventional
renormalization procedure.
\par
The propagator in eq. (\ref{infinite.4}) describes
a complex scalar field. When  one looks at the interaction
part of the action, some appalling features immediately show up:
the vertices  are non-polynomial, they contain
powers of the momentum (up to second power)
and moreover they do not conserve additively the charge
suggested by the free field part. For instance the three legs
vertices are
\begin{eqnarray}
i\Lambda^{D-4}\int d^Dx \Bigl[\partial^\mu \chi^* \partial_\mu \chi
-\frac{1}{2} m^2 \chi^* \chi
\Bigr] \,\,(\chi^*+\chi)
\label{conv.1}
\end{eqnarray}
while the four legs are
\begin{eqnarray}
i\Lambda^{D-4}\int d^Dx \Bigl[\partial^\mu \chi^* \partial_\mu \chi \frac{1}{2}\,\,(\chi^*+\chi)^2
-m^2\Bigl (\frac{1}{6} (\chi^* \chi^3+ \chi^{*3} \chi)
+\frac{1}{4}(\chi^* \chi)^2
\Bigr)
\Bigr]. 
\label{conv.2}
\end{eqnarray}
It is clear that already at one loop the number of independent
divergent amplitudes is infinite. Since standard
renormalization procedure requires that for any divergent 1PI independent
amplitude one must introduce the corresponding local operator
in the classical action, it is clear that the conventional
approach takes to  a dead-end.
\par
In the sequel  I use the method developed for the nonlinear sigma
model, for massive Y-M theories and for the Electroweak Model.
This amount to study the invariance properties of the path-integral
measure, to derive the LFE's associated to the invariance, to establish
the hierarchy among the amplitudes, to fix the number of independent
ancestor amplitudes (via Weak Power Counting (WPC) criterion) and finally
to develop the subtraction strategy for the infinities. Then it is
straightforward to check that the $\phi$-two-point function is that
of a free field.
%
\section{The Complete Set of external Sources}
\label{sec:es}
The path integral measure is
\begin{eqnarray}
\prod_x{\cal D}[\phi(x)]{\cal D}[\phi^*(x)]= \prod_x e^{\chi(x)}
e^{\chi^*(x)}
{\cal D}[\chi(x)]{\cal D}[\chi^*(x)]
\label{exp.3}
\end{eqnarray}
and it is invariant under the local rotations
\begin{eqnarray}&&
\delta_\alpha\phi(x) =i \alpha(x) \phi(x), \qquad \alpha(x) \in {\cal R}
\nonumber\\&&
\delta_\alpha\chi(x)= i\frac{\alpha(x) \phi(x)}{1+\phi(x)}
\label{exp.4}
\end{eqnarray}
and local translations
\begin{eqnarray}&&
\delta_\beta\phi(x) =\beta(x),\qquad \delta_\beta\phi^*(x) =\beta^*(x),\qquad \beta(x) \in {\cal C}
\nonumber\\&&
\delta_\beta\chi(x)= \frac{\beta(x)}{1+\phi(x)},\qquad \delta_\beta\chi^*(x)= \frac{\beta^*(x)}{1+\phi^*(x)}.
\label{exp.5}
\end{eqnarray}
If  one chooses to integrate over the variables $\chi,\,\,\chi^*$,
the transformations in eqs. (\ref{exp.4}) and (\ref{exp.5})
are nonlinear.  One needs a complete set of sources in order
to handle the composite operators intervening in the whole algebra.
By starting with
\begin{eqnarray}
\Lambda^{D-4}\int d^D x [\partial_\mu  \phi^*\partial^\mu  \phi
-m^2 \phi^* \phi + J^*\phi+ J \phi^*]
\label{infinite.4.1}
\end{eqnarray}
and the path integral external field-sources
\begin{eqnarray}\int d^D x [J_0^* \chi   + J_0 \chi^*],
\label{infinite.4.2}
\end{eqnarray}
under $\delta_\alpha$  one needs the extra sources $F^\mu,K$
\begin{eqnarray}
\Lambda^{D-4}\int d^D x [F^\mu(i\phi^*\partial_\mu   \phi-i\partial_\mu   \phi^* \phi)
+K\phi^* \phi].
\label{infinite.5}
\end{eqnarray}
Under $\delta_\beta$  one needs the sources
\begin{eqnarray}
\Lambda^{D-4}\int d^D x [J_1^* \frac{1}{1+\phi(x)}+ J_1 \frac{1}{1+\phi^*(x)}]
\label{infinite.6}
\end{eqnarray}
and then
\begin{eqnarray}&&
\delta_\alpha \frac{1}{(1+\phi(x))^n}= -n \frac{1}{(1+\phi(x))^{(n+1)}}
i\phi\alpha(x)= -n i \frac{1}{(1+\phi(x))^n}\alpha(x)
\nonumber\\&&
 + n i\frac{1}{(1+\phi(x))^{(n+1)}}\alpha(x)
\nonumber\\&&
\delta_\beta \frac{1}{(1+\phi(x))^n}= 
-n \frac{1}{(1+\phi(x))^{(n+1)}}\beta(x).
\label{infinite.7}
\end{eqnarray}
Thus the complete set of sources fixes the effective action at the
tree level
\begin{eqnarray}&&
\Gamma^{(0)}
= \Lambda^{D-4}\int d^D x \Bigl[\partial_\mu  \phi^*\partial^\mu  \phi
-m^2 \phi^* \phi + J^*\phi+ J \phi^*
\nonumber\\&&
+F^\mu \, i(\phi^*\partial_\mu   \phi-\partial_\mu   \phi^* \phi)
+K\phi^* \phi
\nonumber\\&&
+\sum_{n=1}^\infty J_n^*\frac{1}{(1+\phi(x))^n}
+\sum_{n=1}^\infty J_n\frac{1}{(1+\phi^*(x))^n}\Bigr]\,.
\label{infinite.8}
\end{eqnarray}
%

\section{The Local Functional Equation for Rotations}
Since the path integral measure is invariant under local rotations, 
the functional must be invariant under
the change of coordinates (\ref{exp.4}).
By standard procedure, i.e. by expanding in $\alpha$,  one gets
the LFE for the generating functional of
the connected amplitudes
\begin{eqnarray}&&
\Biggl\langle~\partial^\mu i~\Bigl[\phi^*\partial_\mu \phi
-\partial_\mu \phi^*\phi \Bigr]
+i\Bigl[ J^*\phi-J \phi^*\Bigr]
+2\partial^\mu( F_\mu\phi^* \phi )
\nonumber\\&&
+i~\Lambda^{-D+4}\Bigl[ J_0^*\Bigl(   1 
 -\frac{1}{(1+\phi(x))}\Bigr)  - J_0 \Bigl(   1 
 -\frac{1}{(1+\phi^*(x))}\Bigr)\Bigr]
\nonumber\\&&
+\sum_{n=1}^\infty J_n^*\Bigl[  -n i \frac{1}{(1+\phi(x))^n}
 + n i\frac{1}{(1+\phi(x))^{(n+1)}}\Bigr]
\nonumber\\&&
+\sum_{n=1}^\infty J_n\Bigl[  n i \frac{1}{(1+\phi^*(x))^n}
 - n i\frac{1}{(1+\phi^*(x))^{(n+1)}}\Bigr]
\Biggr\rangle~=0\, ,
\label{infinite.10}
\end{eqnarray}
where the brackets denote the weighted mean value over the paths.
It is worth to introduce the notation
\begin{eqnarray}&&
{\cal R}[\alpha]W \equiv \int d^Dx\alpha(x)\Bigl[
\partial^\mu \frac{\delta W}{\delta F^\mu}+i(J^*\frac{\delta W}{\delta J^*}
-J\frac{\delta W}{\delta J})
+2\partial^\mu(F_\mu\frac{\delta W}{\delta K})
\nonumber\\&&
+iJ_0^*(1-\Lambda^{-D+4}\frac{\delta W}{\delta J_1^*})
-iJ_0(1-\Lambda^{-D+4}\frac{\delta W}{\delta J_1})
\nonumber\\&&
+i\sum_{n=1}\, n \,J_n^*\Bigl (- \frac{\delta W}{\delta J_n^* }
+\frac{\delta W}{\delta J_{n+1}^* }  \Bigr)
-i\sum_{n=1}\, n \,J_n\Bigl (- \frac{\delta W}{\delta J_n }
+\frac{\delta W}{\delta J_{n+1} }  \Bigr)\Bigr].
\label{calge.1p}
\end{eqnarray}
Then eq. (\ref{infinite.10}) for the generating functional $W$
(which depends on $F,K,$ $J,J^*,J_0, J_0^*,\{J_n,n=1\cdots\}$)
of the connected amplitudes becomes
\begin{eqnarray}&&
{\cal R}[\alpha]~ W =0.
\label{calge.2p}
\end{eqnarray}
\par

\section{The Local Functional Equation for Translations}
Similarly  one can obtain the LFE for the translations.
The change of coordinates of eq. (\ref{exp.5}) generates
the identity
\footnote{As already stressed in Ref. \cite{Ferrari:2005ii}
the Euler-Lagrange equation for $\chi$
\begin{eqnarray*}&&
(1+\phi)
\Bigl[ -(\Box +m^2) \phi^* + J^*
-2 i F^\mu \partial_\mu   \phi^* - i\partial_\mu  F^\mu \phi^*
+K \phi^*
\nonumber\\&&
   + \Lambda^{-D+4}J_0^* \frac{1}{1+\phi(x)}
-\sum_{n=1}^\infty  n ~J_n^*\frac{1}{(1+\phi(x))^{(n+1)}}
\Bigr] =0
\end{eqnarray*}
and the corresponding Schwinger-Dyson equation are conceptually
different from eqs. (\ref{infinite.12p}) and (\ref{calge.23p}).
}
\begin{eqnarray}&&
\Biggl\langle -(\Box +m^2) \phi^* + J^*
-2 i F^\mu \partial_\mu   \phi^* - i\partial_\mu  F^\mu \phi^*
+K \phi^*
\nonumber\\&&
   + \Lambda^{-D+4}J_0^* \frac{1}{1+\phi(x)}
-\sum_{n=1}^\infty  n ~J_n^*\frac{1}{(1+\phi(x))^{(n+1)}}
\Biggr\rangle =0.
\label{infinite.12p}
\end{eqnarray}
 As before the operator is defined
\begin{eqnarray}&& \!\!\!
{\cal T}[\beta] W\equiv \int d^Dx\beta(x)\Bigl[
-(\Box +m^2)\frac{\delta W}{\delta J}+J^*
-2 i F^\mu \partial_\mu \frac{\delta W}{\delta J}
\nonumber\\&&
- i\partial_\mu  F^\mu\frac{\delta W}{\delta J}
+K\frac{\delta W}{\delta J}+\Lambda^{-D+4}J_0^*\frac{\delta W}{\delta J_1^*}
-\sum_{n=1}\, n \,J_n^*\frac{\delta W}{\delta J_{n+1}^*}
\Bigr].
\label{calge.13p}
\end{eqnarray}
Eq. (\ref{infinite.12p}) becomes
\begin{eqnarray}&&
{\cal T}[\beta]~ W =0.
\label{calge.23p}
\end{eqnarray}
It is important to establish the algebra of ${\cal R},\, \, {\cal T}$.
By a straightforward calculation 
\begin{eqnarray}&&
\Bigl[{\cal R}[\alpha],{\cal T}[\beta]\Bigr]
=-i {\cal T}[\alpha\beta].
\label{calge.24p}
\end{eqnarray}
It should be noticed that both eqs. (\ref{calge.2p}) and  (\ref{calge.23p})
are not aware of the choice  one might operate for the integration variables:
either $\phi,\phi^*$ or $\chi,\chi^*$.
Moreover it is worth noticing that both equations are linear in $W$.

\section{Effective Action Functional}
\label{sec:eff}
The situation becomes rather interesting when we
derive the effective action functional $\Gamma$, via Legendre 
transformations. This
step is necessary in order to set up a strategy for
the subtraction of the infinities of the perturbative
expansion. The aim of this work is to formulate  the field theory 
in terms of polar coordinates, then the Legendre transformation 
is done on the 
variables $\chi,\chi^*$. $\Gamma$ obeys  the following LFE's: for the rotations
\begin{eqnarray}&&
\partial^\mu \frac{\delta\Gamma}{\delta F^\mu}
+i\Bigl[ J^*\frac{\delta\Gamma}{\delta J^*}
-J \frac{\delta\Gamma}{\delta J}\Bigr]
+2  \partial^\mu( F_\mu \frac{\delta\Gamma}{\delta K})
\nonumber\\&&
 -i\frac{\delta\Gamma}{\delta \chi}\Bigl(   1 
 -\Lambda^{-D+4}\frac{\delta\Gamma}{\delta J_1^*}\Bigr)
+i \frac{\delta\Gamma}{\delta \chi^*}\Bigl(   1 
 -\Lambda^{-D+4}\frac{\delta\Gamma}{\delta J_1}\Bigr)
\nonumber\\&&
+\sum_{n=1}^\infty n~i
\Biggl( J_n^*\Bigl[  - \frac{\delta\Gamma}{\delta J_n^*}
 + \frac{\delta\Gamma}{\delta J_{(n+1)}^*}\Bigr]
+J_n~\Bigl[ \frac{\delta\Gamma}{\delta J_n}
 - \frac{\delta\Gamma}{\delta J_{(n+1)}}\Bigr]
\Biggr)
~=0
\label{infinite.11.1}
\end{eqnarray}
and for the translations
\begin{eqnarray}&&
-\Bigl(\Box 
+m^2 \Bigr)\frac{\delta\Gamma}{\delta J} + \Lambda^{D-4}J^*
-2 i F^\mu \partial_\mu  \frac{\delta\Gamma}{\delta J}  
- i\partial_\mu  F^\mu \frac{\delta\Gamma}{\delta J}
+K \frac{\delta\Gamma}{\delta J}
\nonumber\\&&
   -\Lambda^{-D+4}\frac{\delta\Gamma}{\delta \chi} \frac{\delta\Gamma}{\delta J_1^*}
-\sum_{n=1}^\infty  n ~J_n^*\frac{\delta\Gamma}{\delta J_{(n+1)}^*} =0.
\label{infinite.14p.1}
\end{eqnarray}
By using these equations and the tree-level effective action
$\Gamma^{(0)}$ in eq.
(\ref{infinite.8}) it is possible to reconstruct the perturbative
series in powers of $\hbar$ (loop-expansion). 
\subsection{Hierarchy}
Eqs. (\ref{infinite.11.1}) and (\ref{infinite.14p.1})
guarantee full hierarchy: every amplitude with at least
one $\chi$ or $\chi^*$ leg (descendant) can be obtained
from those (ancestors) without any of them (the elementary fields)
\cite{Ferrari:2005ii}, \cite{Bettinelli:2007zn}.
This is a great advantage since the number of independent
counterterms for the ancestors is finite at every order 
in the loop expansion, as it will be discussed in Sections 
\ref{sec:wpc} and  \ref{sec:sol}.
\par
Hierarchy is  here illustrated by an explicit example. By
taking the derivative of (\ref{infinite.14p.1}) with
respect to $J^*(y)$ and by putting all sources and fields to zero
one gets
\begin{eqnarray}&&
-\Bigl(\Box +m^2 \Bigr)\frac{\delta^2\Gamma}{\delta J(x)\delta J^*(y)} + 
\Lambda^{D-4}\delta(x-y)
-\Lambda^{-D+4}\frac{\delta^2\Gamma}{\delta \chi(x)\delta J^*(y)} \frac{\delta\Gamma}{\delta J_1^*} 
=0.
\nonumber\\&&
\label{eff.11}
\end{eqnarray}
Thus the two-point function $\chi-J^*$ is known in terms
of the two-point function $J-J^*$ and of the one-point $J_1^*$.
In a perturbative approach  one starts from $\Gamma^{(0)}$
which is a solution of both equations (\ref{infinite.11.1})
and (\ref{infinite.14p.1}), by construction. Thus eq. (\ref{eff.11})
is realized at the tree level since
\begin{eqnarray}&&
\frac{\delta^2\Gamma^{(0)}}{\delta J(x)\delta J^*(y)}=0
\nonumber\\&&
\frac{\delta^2\Gamma^{(0)}}{\delta \chi(x)\delta J^*(y)}= \Lambda^{D-4}
\delta(x-y)
\nonumber\\&&
\frac{\delta\Gamma^{(0)}}{\delta J_1^*} =\Lambda^{D-4} \, .
\label{eff.12}
\end{eqnarray}
At the one-loop level one gets from eq. (\ref{eff.11})
\begin{eqnarray}
\Lambda^{-D+4}
\frac{\delta\Gamma^{(0)}}{\delta J_1^*}  \frac{\delta^2\Gamma^{(1)}}{\delta \chi(x)\delta J^*(y)}=
-\Bigl(\Box +m^2 \Bigr)\frac{\delta^2\Gamma^{(1)}}{\delta J(x)\delta J^*(y)}.
\label{eff.13}
\end{eqnarray}
In $D$ dimensions and by using the vertices in eqs.
(\ref{conv.1}) and (\ref{conv.2}) the relevant quantities 
are
\begin{eqnarray}&&
\frac{\delta^2\Gamma^{(1)}}{\delta J(x)\delta J^*(y)}=
\frac{i}{2} [\Delta_m(x-y)]^2
\nonumber\\&&
\frac{\delta^2\Gamma^{(1)}}{\delta \chi(x)\delta J^*(y)}= 
- \frac{i}{2} (\Box+m^2)[\Delta_m(x-y)]^2
\label{eff.14}
\end{eqnarray}
where the free $\chi$ propagator is
\begin{eqnarray}
\Delta_m(x-y)\equiv \Lambda^{D-4}\langle 0|T(\chi(x)\chi^*(y))|0\rangle.
\label{eff.15}
\end{eqnarray}
The derivative of the complex conjugate of eq. (\ref{infinite.14p.1}) 
with respect to $\chi$ yields a further descendant amplitude
\begin{eqnarray}
\Bigl(\Box +m^2 \Bigr)\frac{\delta^2\Gamma}{\delta J^*(x)\delta \chi(y)} 
+\Lambda^{-D+4}\frac{\delta^2\Gamma}{\delta \chi^*(x)\delta \chi(y)} 
\frac{\delta\Gamma}{\delta J_1} 
=0.
\label{eff.15.1}
\end{eqnarray}
If more $\chi$ and  $\chi^*$ insertions are needed, further $J_n, J_m^*$ will intervene in increasing number.
Of course one is very much  interested to know if the two point function
(connected) turns out correct in this formalism. Indeed one verifies that
at one loop the necessary cancellation occurs and
\begin{eqnarray}&&
\frac{\delta^2 W^{(0)}}{\delta J^*(x)\delta J(y)}= \Delta_m(x-y)
\nonumber\\&&
\frac{\delta^2 W^{(1)}}{\delta J^*(x)\delta J(y)}= 0,
\label{eff.16}
\end{eqnarray}
i.e. $\phi$ remains a free field. A further point of interest
is whether the theory makes any sense at $D=4$.
\section{$D=4$ Limit}
\label{sec:lim}
The fundamental question is whether one can define
a sensible theory at $D=4$. If one succeeds then
the existence of field-coordinate transformation is
proven by construction. In this Section I use mostly heuristic
arguments: the proofs have been given elsewhere
\cite{Bettinelli:2007zn}
and moreover the main points should not be masked by too many
details. Let 
\begin{eqnarray}
\hat\Gamma \equiv \Gamma^{(0)}+ \sum_{n=1}^\infty \hat\Gamma^{(n)}
\label{eff.17}
\end{eqnarray}
where $\hat\Gamma^{(n)}$ are the local counterterms, that will be
constructed on the ongoing. The generating functional of the Feynman amplitudes is given by
\begin{eqnarray}
Z=\exp iW \simeq \int e^{\chi+\chi^*}{\cal D} [\chi^*]{\cal D}[\chi]\,\,\,
{\Large e^{i\hat\Gamma}}\,\,\,\exp \Bigl(i\int d^D x (\chi J_0^*+\chi^* J_0) 
\Bigr)\, .
\label{eff.18.0}
\end{eqnarray}
By using the linear operator of eq. (\ref{calge.1p}) for the rotations
one gets
\begin{eqnarray}&&
{\cal R}[\alpha]Z
=\Biggl[
\partial^\mu \frac{\delta\hat\Gamma}{\delta F^\mu}
+i\Bigl[ J^*\frac{\delta\hat\Gamma}{\delta J^*}
-J \frac{\delta\hat\Gamma}{\delta J}\Bigr]
+2  \partial^\mu( F_\mu \frac{\delta\hat\Gamma}{\delta K})
\nonumber\\&&
 -i\frac{\delta\hat\Gamma}{\delta \chi}\Bigl(   1 
 -\Lambda^{-D+4}\frac{\delta\hat\Gamma}{\delta J_1^*}\Bigr)
+i \frac{\delta\hat\Gamma}{\delta \chi^*}\Bigl(   1 
 -\Lambda^{-D+4}\frac{\delta\hat\Gamma}{\delta J_1}\Bigr)
\nonumber\\&&
+\sum_{n=1}^\infty n~i
\Biggl( J_n^*\Bigl[  - \frac{\delta\hat\Gamma}{\delta J_n^*}
 + \frac{\delta\hat\Gamma}{\delta J_{(n+1)}^*}\Bigr]
+J_n~\Bigl[ \frac{\delta\hat\Gamma}{\delta J_n}
 - \frac{\delta\hat\Gamma}{\delta J_{(n+1)}}\Bigr]
\Biggr)
\Biggr]\cdot Z=0,
\nonumber\\&&
\label{eff.18}
\end{eqnarray}
where the dot indicates the insertion of the operator.
Eq. (\ref{eff.18}) states that if $\hat\Gamma$ (the counterterms) obeys the same equation (\ref{infinite.11.1}) as the effective action, then the LFE is valid
for the generating functionals.
Similar result is valid for the translations in eq. (\ref{calge.13p}).
\begin{eqnarray}&& \!\!\!
{\cal T}[\beta] Z
=\Biggl[
-\Bigl(\Box 
+m^2 \Bigr)\frac{\delta\hat\Gamma}{\delta J} + \Lambda^{D-4}J^*
-2 i F^\mu \partial_\mu  \frac{\delta\hat\Gamma}{\delta J}  
- i\partial_\mu  F^\mu \frac{\delta\hat\Gamma}{\delta J}
\nonumber\\&&
+K \frac{\delta\hat\Gamma}{\delta J}
   -\Lambda^{-D+4}\frac{\delta\hat\Gamma}{\delta \chi} \frac{\delta\hat\Gamma}{\delta J_1^*}
-\sum_{n=1}^\infty  n ~J_n^*\frac{\delta\hat\Gamma}{\delta J_{(n+1)}^*}
\Biggr]\cdot Z=0.
\label{eff.19}
\end{eqnarray}
Thus also for the translations, if $\hat\Gamma$ satisfies the same equation
as the effective action functional (\ref{infinite.14p.1}), then the LFE is
valid for the generating functionals.
\par
 {
Eqs. (\ref{eff.18}) and (\ref{eff.19}) show that the perturbative series
in $D$ dimensions (without counterterms) satisfies both LFE's. 
The problem now is to show that
after the subtractions and the limit $D=4$ the resulting finite
theory still satisfies the same equations, that the subtraction procedure
is achieved by local counterterms and that the two-point-function
of the $\phi$-fields is that of a free theory.
}
%
\subsection{Linearized Operators}
The discussion of the perturbative construction of the coordinate-field
transformations makes use of the linearized form of the operators
acting on $\Gamma$ as in eqs. (\ref{infinite.11.1}) and (\ref{infinite.14p.1}).
Rotations:
\begin{eqnarray}&&
\rho(x)\equiv
-\partial^\mu\frac{\delta}{\delta F^\mu}
-i\Bigl[ J^*\frac{\delta}{\delta J^*}
-J \frac{\delta}{\delta J}\Bigr]
-2  \partial^\mu\Bigl( F_\mu \frac{\delta}{\delta K}\Bigr)
\nonumber\\&&
 -i\frac{\delta\Gamma^{(0)}}{\delta \chi}\frac{\delta}{\delta J_1^*}
+i \frac{\delta\Gamma^{(0)}}{\delta \chi^*}\frac{\delta}{\delta J_1}
+ i\Bigl(   1 
 -\frac{\delta\Gamma^{(0)}}{\delta J_1^*}\Bigr)\frac{\delta}{\delta \chi}
-i \Bigl(   1 
 -\frac{\delta\Gamma^{(0)}}{\delta J_1}\Bigr)\frac{\delta}{\delta \chi^*}
\nonumber\\&&
-i\sum_{n=1}^\infty n
\Biggl( J_n^*\Bigl[  - \frac{\delta}{\delta J_n^*}
 + \frac{\delta}{\delta J_{(n+1)}^*}\Bigr]
+J_n~\Bigl[ \frac{\delta}{\delta J_n}
 - \frac{\delta}{\delta J_{(n+1)}}\Bigr]
\Biggr)
\nonumber\\&&
\label{bra.1}
\end{eqnarray}
and translations:
\begin{eqnarray}&&
\tau(x)\equiv
\Bigl(\Box 
+m^2 \Bigr)\frac{\delta}{\delta J} 
+i2 F^\mu \partial_\mu  \frac{\delta}{\delta J} 
+i\partial_\mu  F^\mu \frac{\delta}{\delta J}
\nonumber\\&&
-K \frac{\delta}{\delta J}
   +\frac{\delta\Gamma^{(0)}}{\delta \chi} \frac{\delta}{\delta J_1^*}
   +\frac{\delta\Gamma^{(0)}}{\delta J_1^*}\frac{\delta}{\delta \chi} 
+\sum_{n=1}^\infty  n ~J_n^*\frac{\delta}{\delta J_{(n+1)}^*}.
\label{bra.3}
\end{eqnarray}
%
%
\subsection{Subtraction Strategy}
The subtraction is performed by iteration, according to 
the forest formula \cite{subtraction}. The divergent part of the effective
action is given by the pole part in the Laurent expansion
of 
\begin{eqnarray}
\frac{1}{\Lambda^{D-4}} \Gamma^{(n)}\Bigr|_{\rm Pole \,\,\, Part}.
\label{eff.20}
\end{eqnarray}
The divergent part is removed by adding a counterterm to the effective
action, i.e. the result of the eq. (\ref{eff.20}) is written as 
local counterterm in terms of variables in $D$ dimensions and thus
 $\hat \Gamma^{(n)}$ is generated. This 
procedure can be written in a symbolic way by
\begin{eqnarray}
\hat\Gamma^{(n)} = \Lambda^{D-4} \Bigg(-
\frac{1}{\Lambda^{D-4}} \Gamma^{(n)}\Biggr)\Bigr|_{\rm Pole \,\,\, Part}.
\label{eff.21}
\end{eqnarray}
The subtraction strategy stands on two important results.
\par\noindent
{\Large\bf Proposition1}. Let the counterterms in $\hat\Gamma$
satisfy the eqs. (\ref{infinite.11.1}) and (\ref{infinite.14p.1})
then the subtraction up to order $n-1$ breaks the LFE's for
the effective action $\Gamma$ at order $n$ by the local terms:
\begin{eqnarray}&&
\rho \Gamma^{(n)}
 {-}i\Lambda^{-D+4}\sum_{j=1}^{n-1}
\Bigl(\frac{\delta\Gamma^{(n-j)}}{\delta \chi}
\frac{\delta\Gamma^{(j)}}{\delta J_1^*}
-
\frac{\delta\Gamma^{(n-j)}}{\delta \chi^*}
\frac{\delta\Gamma^{(j)}}{\delta J_1}
\Bigr)
\nonumber\\&&
=
 {-}
i\Lambda^{-D+4}\sum_{j=1}^{n-1}
\Bigl(\frac{\delta\hat\Gamma^{(n-j)}}{\delta \chi}
\frac{\delta\hat\Gamma^{(j)}}{\delta J_1^*}
-
\frac{\delta\hat\Gamma^{(n-j)}}{\delta \chi^*}
\frac{\delta\hat\Gamma^{(j)}}{\delta J_1}
\Bigr)
\label{infinite.11.2}
\end{eqnarray}
and for the translations,
\begin{eqnarray}&&
\tau \Gamma^{(n)}+\Lambda^{-D+4}\sum_{j=1}^{n-1}
\frac{\delta\Gamma^{(n-j)}}{\delta \chi}
\frac{\delta\Gamma^{(j)}}{\delta J_1^*}
=
\Lambda^{-D+4}\sum_{j=1}^{n-1}
\frac{\delta\hat\Gamma^{(n-j)}}{\delta \chi}
\frac{\delta\hat\Gamma^{(j)}}{\delta J_1^*}.
\label{infinite.14p.2}
\end{eqnarray}
It should be stressed that $\Gamma^{(n)}$ can diverge
in the limit $D=4$ since the relevant poles have not been
subtracted, while on the same ground $\Gamma^{(n-j)}, j>0,$
is finite. 
\par\noindent
A heuristic proof of the Proposition 1 goes as follows.
$\hat\Gamma$ satisfies the LFE's by assumption.
Then if one adds   $\rho~\hat\Gamma^{(n)}$ and $\tau~\hat\Gamma^{(n)}$
 to both sides
of   eqs. (\ref{infinite.11.2}) and (\ref{infinite.14p.2})
respectively, 
then the LFE's become valid at order $n$ because of equations
(\ref{eff.18}) and  (\ref{eff.19}).
\par\noindent 
A proof can be given also by using the grading
of $\Gamma^{(n)}$ according to the total power of $\hbar$
of the counterterms \cite{Bettinelli:2007zn}.
\par
The second results can be derived directly from Proposition 1.
\par\noindent
{\Large\bf Proposition2}
The subtraction rules consisting in the removal of
the sole pole part in the Laurent expansion around $D=4$
of the effective action $\Gamma $ (eq. (\ref{eff.21}))
yields a  $\hat\Gamma $ that obeys the LFE's. The counterterms
are local.
\par
According to eq. (\ref{eff.21}) the multiplication by
$\Lambda^{-D+4}$ of both sides of eqs. (\ref{infinite.11.2}) 
and (\ref{infinite.14p.2}) reduces the right hands sides to
pure pole terms (no finite parts). Thus the removal of
the sole pole part from $\Lambda^{-D+4}~\Gamma$ reestablishes
the validity  of the LFE's.
\par
The whole subtraction procedure is invalidated by
any finite renormalization, e.g. by on-shell renormalization.
In fact it is true that,  at a given order of the perturbation expansion,
the equations (\ref{infinite.11.2}) 
and (\ref{infinite.14p.2}) are still valid if one adds to
the  counterterm $\hat\Gamma^{(n)}$ any local solution ${\cal M}$
of the homogeneous equations 
\begin{eqnarray}&&
\rho {\cal M}=0
\nonumber\\&&
\tau {\cal M}=0.
\label{eff.22}
\end{eqnarray}
In doing so, however, the pole structure of the breaking
terms in eqs. (\ref{infinite.11.2}) 
and (\ref{infinite.14p.2}) is modified by finite (at $D=4$) operators. 
Consequently
there is no more a strategy (e.g. the pure pole subtraction) 
to avoid the increasing of
the number of free parameters and  the breakdown of the
perturbative expansion. This unwanted feature of finite renormalizations
does not appear in some special cases as, for instance, in a rescaling
of $\Lambda$ (e.g. $\overline{\rm MS}$).
\section{Weak Power Counting}
\label{sec:wpc}
The LFE's provide full hierarchy, as discussed 
in Section \ref{sec:eff}. In particular all the 
divergent amplitudes involving the elementary fields 
$\chi~ \chi^*$ are organized and controlled by a finite
number of divergent ancestor amplitudes at any given order
of the perturbative expansion. To establish which
are the relevant ancestor amplitudes, the  WPC criterion
is very useful. 
\par
The degree of divergence of a graph $G$
for an ancestor amplitude can be evaluated in the
usual fashion. Let $I$ be the number of internal $\chi$ propagators,
$N_F$   the number of external $F_\mu$ legs and 
 $N_X$ those of $X\in\{K, J,J^*, J_i, J_i^* | i=1,\cdots\}$.
$V_{jk}$ denotes the number of vertices with $k$ $\chi$-lines 
and $j$ derivatives and $N_{Xk}$ the number of $X$ external
sources with $k$ internal lines of $\chi$'s attached.
The superficial degree of divergence $\delta( G)$
for a graph can be bounded by using standard arguments.
One has ($n$ number of loops)
\begin{eqnarray}&&
\delta(G)= D~ n -2I +\sum_{j,k}j~V_{jk}+ N_{F}
\nonumber\\&&
n=I-\sum_{j,k} V_{jk}-N_F-\sum_X N_X+1 
\label{wpc.2}
\end{eqnarray}
By removing $I$ from these  two equations one gets
\begin{eqnarray}
\delta(G)= D~ n-2n  -\sum_{j,k}(2-j)~V_{jk}- N_{F}
-2 \sum_X N_X+2.
\label{wpc.3}
\end{eqnarray}
The classical action (\ref{infinite.4}) has vertices with $j\leq 2$,
therefore it can be stated that
\begin{eqnarray}
\delta(G)\leq  n(D-2)+2 - N_{F}-2 \sum_X N_X.
\label{wpc.4}
\end{eqnarray}
By arguments similar to those used for  eq. (\ref{wpc.4}),
one can prove that the subtraction procedure indicated
by eq. (\ref{eff.22}) does not modify the upper bound on the
superficial divergence (WPC theorem).
\par
Eq. (\ref{wpc.4}) restricts the set of divergent ancestor amplitudes
to those with a finite number of external legs.
It will be shown, in Section \ref{sec:sol}, that
the number of counterterms becomes finite after the analysis of
the local solutions of eqs (\ref{infinite.11.2}) and
(\ref{infinite.14p.2}). 
\par
At one loop and $D=4$ the possible divergent ancestor
amplitudes are: $K$ one-point-functions,
$F_\mu-F_\nu$, $K-F_\mu$, $X-X'$
two-point-functions, $(F_\mu-F_\nu)-X$
three-point-functions and $F_\mu$ four-point-functions.

\section{Local Solutions of the Linear LFE's}
\label{sec:sol}
The discussion on the divergent ancestor amplitudes
has shown that, at any order of the perturbative
expansion, the number of external legs must
be finite. However the number of external sources
is infinite, thus the constraint given by the
WPC criterion is not enough to make finite the number of 
counterterms. In this Section other constraints on the
counterterms are found. These constraints limit the
counterterms to a finite number of local invariant operators.
\par
In the present Section  the local solutions
of the homogeneous eqs. (\ref{eff.22}) are considered, since the aim is to illustrate
the method only at the one loop level. 
If higher loop are of interest, then it is necessary to solve the
following
nonlinear equations and  an example is briefly
discussed in Section \ref{sec:hlc}. For the rotations
\begin{eqnarray}&&
\rho \hat\Gamma^{(n)}
=
i\Lambda^{-D+4}\sum_{j=1}^{n-1}
\Bigl(\frac{\delta\hat\Gamma^{(n-j)}}{\delta \chi}
\frac{\delta\hat\Gamma^{(j)}}{\delta J_1^*}
-
\frac{\delta\hat\Gamma^{(n-j)}}{\delta \chi^*}
\frac{\delta\hat\Gamma^{(j)}}{\delta J_1}
\Bigr)
\label{infinite.11.3}
\end{eqnarray}
and for the translations
\begin{eqnarray}
\tau \hat\Gamma^{(n)}
=
-\Lambda^{-D+4}\sum_{j=1}^{n-1}
\frac{\delta\hat\Gamma^{(n-j)}}{\delta \chi}
\frac{\delta\hat\Gamma^{(j)}}{\delta J_1^*}.
\label{infinite.14p.3}
\end{eqnarray}
\par
In order to find the relevant local solutions of the linearized
LFE's one looks for suitable variables that have simple
transformations under $\rho$ and $\tau$. This method is very
similar to the bleaching technique introduced in Ref. 
\cite{analysis}.
Moreover there are some helping directions in the search of local invariants 
at one loop. For instance they need to be at most quadratic in the sources $X$ 
and quartic in $F_\mu$. 
\par
The following notation is useful 
\begin{eqnarray}&&
\rho[\alpha]\equiv \int d^Dx \alpha(x)\rho(x)
\nonumber\\&&
\tau[\beta]\equiv \int d^Dx \beta(x)\tau(x)
\label{sou.0}
\end{eqnarray}
\begin{eqnarray}&&
{\cal S}[\phi,\phi^*, F,K]\equiv
\int d^D z  \Bigl(\partial_\mu  \phi^*\partial^\mu  \phi
-m^2 \phi^* \phi 
 {+J^*\phi+ J\phi^*}
\nonumber\\&&
+F^\mu i\phi^*\su{\leftrightarrow}{\partial}_\mu   \phi
+K\phi^* \phi\Bigr).
\label{lin.5}
\end{eqnarray}
One gets
 {
\begin{eqnarray}&&
-J_0^*\equiv\frac{\delta\Gamma^{(0)}}{\delta \chi}
=(1+\phi)\Bigl[\frac{\delta {\cal S}}{\delta \phi}-\sum_{n=1} 
\frac{nJ_n^*}{(1+\phi)^{n+1}}
\Bigr]
\nonumber\\&&
=(1+\phi)\frac{\delta {\cal S}}{\delta \phi}-\sum_{n=1} \frac{nJ_n^*}{(1+\phi)^{n}}.
\label{lin.5.1}
\end{eqnarray}
}
The rotation transformations generated by $\rho$ 
in eq. (\ref{bra.1}) are
\begin{eqnarray}
\begin{array}{ll}
\rho[\alpha] \chi = i \alpha \frac{\phi}{1+\phi} &
\rho[\alpha]\phi  = i \alpha \phi\\
\rho[\alpha] J=   i \alpha J &
\rho[\alpha] F^\mu  = \partial^\mu  \alpha \\
\rho[\alpha] K  = 2 F^\mu \partial_\mu  \alpha &
\rho[\alpha] J_1
= -  i \alpha\Bigl[J_1- \frac{\delta\Gamma^{(0)}}{\delta \chi^*}\Bigr]
\\
\rho[\alpha] J_n=  -i \alpha(n J_n & \!\!\! \! -( n-1) J_{n-1}) \quad n>1
\end{array}
\label{lin.4.0p}
\end{eqnarray}
and the translations generated by $\tau[\beta]$ in eq. (\ref{bra.3}) are
\begin{eqnarray}
\begin{array}{ll}
\tau[\beta]  \chi =  \beta \frac{1}{1+\phi} &
\tau[\beta] \phi  =  \beta \\
\tau[\beta]  K  =0 &
\tau[\beta]  F^\mu  = 0\\
\tau[\beta]  J=  - \int d^Dy\beta(y)\frac{\delta^2{\cal S}}{\delta \phi (y)\phi^* }
&
\tau[\beta]  J_1^*=   \beta \frac{\delta\Gamma^{(0)}}{\delta \chi}\\
\tau[\beta]  J_n^*= ( n-1) \beta J_{n-1}^*, & n>1 \, . 
\end{array}
\label{lin.4.1p}
\end{eqnarray}
%

%
\subsection{Properties of the rotations generator $\rho[\alpha]$}
The local invariants under rotations are first investigated.
$F_\mu$ plays the r\^ole of abelian gauge fields, thus it is convenient
to define a covariant derivative
\begin{eqnarray}
D_\mu \equiv \partial_\mu -i F_\mu.
\label{cov.1}
\end{eqnarray}
Consider the following quantity
\begin{eqnarray}&&
\rho[\alpha] \frac{\delta {\cal S}}{\delta\phi^* (x)}
=\rho[\alpha] \Bigl[-(\Box+m^2)\phi 
 {+J}+iF_\mu \partial^\mu\phi
+i\partial^\mu(F_\mu\phi ) +K\phi
\Bigr]
\nonumber\\&&
=-i(\Box+m^2)(\alpha\phi)  {+i\alpha J}-F_\mu \partial^\mu(\alpha\phi)
+i\partial_\mu\alpha \partial^\mu\phi
-\partial^\mu(F_\mu\alpha\phi )
\nonumber\\&&
+i\partial^\mu(\partial^\mu\alpha\phi )
 +iK\alpha\phi +2F_\mu \partial^\mu\alpha\phi
\nonumber\\&&
=i \alpha \frac{\delta {\cal S}}{\delta\phi^* (x)}.
\label{al.2}
\end{eqnarray}
One obtains similarly
\begin{eqnarray}
\rho[\alpha] \frac{\delta {\cal S}}{\delta\phi (x)}
= - i \alpha \frac{\delta {\cal S}}{\delta\phi (x)}.
\label{al.2.1}
\end{eqnarray}
From eqs. (\ref{al.2}) and (\ref{al.2.1}) one gets
\begin{eqnarray}&&
\rho[\alpha] \frac{\delta \Gamma^{(0)}}{\delta\chi (x)}
= \rho[\alpha] \frac{\delta {\cal S}}{\delta\chi (x)}
-\rho[\alpha] \sum_{n=1}n \frac{J_n^*}{(1+\phi)^n}
=\rho[\alpha](1+\phi ) \frac{\delta {\cal S}}{\delta\phi (x)}
\nonumber\\&&
-i\alpha \Bigl[J_1^*- \frac{\delta\Gamma^{(0)}}{\delta \chi(x)}\Bigr]
\frac{1}{(1+\phi)}
 + i \alpha(x)\frac{ J_1^*\phi}{(1+\phi)^2}
\nonumber\\&&
-\sum_{n=2}\Bigl[ i \alpha(x)(  n J_n^*(x)- ( n-1) J_{n-1}^*(x))\frac{n}{(1+\phi)^n}
- i \alpha(x)\frac{n^2 J_n\phi}{(1+\phi)^{n+1}}
\Bigr]
\nonumber\\&&
=\rho[\alpha] \frac{\delta {\cal S}}{\delta\phi (x)}
+i\alpha \frac{\delta \Gamma^{(0)}}{\delta\phi (x)}
 -i \alpha(x)\frac{ J_1^*}{(1+\phi)^2}
+i2\alpha(x)\frac{ J_1^*}{(1+\phi)^2}
\nonumber\\&&
- i \alpha(x)\sum_{n=2}\Bigl[   J_n^*(x)\frac{n^2}{(1+\phi)^n}
-  J_{n}^*(x)\frac{n(n+1)}{(1+\phi)^{n+1}}
- \frac{n^2 J_n^*\phi}{(1+\phi)^{n+1}}\Bigr]
\nonumber\\&&
= -i \alpha(x)\frac{\delta {\cal S}}{\delta\phi (x)}
+i\alpha \frac{\delta \Gamma^{(0)}}{\delta\phi (x)}
 +i \alpha(x)\frac{ J_1^*}{(1+\phi)^2}
+i\alpha(x) \sum_{n=2}J_{n}^*(x)\frac{n}{(1+\phi)^{n+1}}
\nonumber\\&&
=-i \alpha(x)\frac{\delta\Gamma^{(0)}}{\delta\phi (x)}
+i\alpha \frac{\delta \Gamma^{(0)}}{\delta\phi (x)}
=0.
\label{al.3}
\end{eqnarray}
and
\begin{eqnarray}
\rho[\alpha] \frac{\delta \Gamma^{(0)}}{\delta\chi^* (x)}
=0.
\label{al.3.1}
\end{eqnarray}
Another interesting local operator is given by the following expression 
\begin{eqnarray}
{\cal N}_0\equiv \Bigl[J_0^*\ln(1+\phi) +\sum_{n=1}
J_n^*\frac{1}{(1+\phi)^n}  \Bigr]\Bigr|_{J_0^*=-\frac{\delta \Gamma^{(0)}}{\delta\chi (x)}}.
\label{phi.1.3p}
\end{eqnarray}
By direct computation one gets
\begin{eqnarray}&&
\rho[\alpha] {\cal  N}_0=
\rho[\alpha]  J_0^*\ln(1+\phi) + i\alpha \frac{J_0^*\phi}{(1+\phi)}
+ i\alpha \frac{J_0^*+J_1^*}{(1+\phi)}
\nonumber\\&&
-i\alpha \frac{J_1^*\phi}{(1+\phi)^2}
- i\alpha\sum_{n=2}n \frac{J_n^*\phi}{(1+\phi)^{n+1}}
+ i\alpha\sum_{n=2}n \frac{J_n^*\phi}{(1+\phi)^n}
- i\alpha\sum_{n=1}n \frac{J_n^*}{(1+\phi)^{n+1}}
\nonumber\\&&
=
\rho[\alpha]  J_0^*\ln(1+\phi) + i\alpha \frac{J_0^*\phi}{(1+\phi)}
+ i\alpha \frac{J_0^*}{(1+\phi)}
\nonumber\\&&
- i\alpha\sum_{n=1}n \frac{J_n^*\phi}{(1+\phi)^{n+1}}
+ i\alpha\sum_{n=1}n \frac{J_n^*\phi}{(1+\phi)^n}
- i\alpha\sum_{n=1}n \frac{J_n^*}{(1+\phi)^{n+1}}
\nonumber\\&&
=
\rho[\alpha]  J_0^*\ln(1+\phi) + i\alpha  J_0^*.
\label{al.3.2}
\end{eqnarray}
By using eq. (\ref{al.3}) one gets
\begin{eqnarray}
\rho[\alpha] {\cal N}_0
=-i\alpha \frac{\delta \Gamma^{(0)}}{\delta\chi (x)}
\label{al.3.3}
\end{eqnarray}
\begin{eqnarray}
\rho[\alpha] {\cal N}_0^*
=i \alpha \frac{\delta \Gamma^{(0)}}{\delta\chi^* (x)}.
\label{al.3.4}
\end{eqnarray}
One can also consider
\begin{eqnarray}&&
{\cal N}_{1}\equiv \Bigl[J_0^*\frac{1}{(1+\phi)}
 -\sum_{n=1}
\frac{nJ_n^*}{(1+\phi)^{n+1}} \Bigr]\Bigr|_{J_0^*=-\frac{\delta \Gamma^{(0)}}{\delta\chi (x)}}
\nonumber\\&&
 {
=-\Bigl[\frac{\delta {\cal S}}{\delta \phi}-\sum_{n=1} 
\frac{nJ_n^*}{(1+\phi)^{n+1}}
\Bigr] -\sum_{n=1}
\frac{nJ_n^*}{(1+\phi)^{n+1}} = -\frac{\delta {\cal S}}{\delta \phi}.
}
\label{phi.1.3pp}
\end{eqnarray}
From eq. (\ref{al.2.1}) one gets
\begin{eqnarray}
\rho[\alpha] {\cal N}_{1}
= -i\alpha {\cal N}_{1}.
\label{al.4}
\end{eqnarray}
Similarly one has ($j>0$)
\begin{eqnarray}&&
{\cal N}_{j}\equiv \Bigl[-J_0^*\frac{(j-1)!}{(1+\phi)^j}
+\sum_{n=1}n(n+1)\cdots(n+j-1)
J_n^*\frac{1}{(1+\phi)^{n+j}} \Bigr]\Bigr|_{J_0^*=-\frac{\delta \Gamma^{(0)}}{\delta\chi (x)}}
\nonumber\\&&
 {
=\Bigl[\frac{(j-1)!}{(1+\phi)^{j-1}}\frac{\delta {\cal S}}{\delta \phi}-\sum_{n=1} 
\frac{n(j-1)! \,J_n^*}{(1+\phi)^{n+j}}
\Bigr] +\sum_{n=1}
\frac{n(n+1)\cdots(n+j-1)J_n^*}{(1+\phi)^{n+j}}
}
\nonumber\\&&
 {
 = \frac{(j-1)!}{(1+\phi)^{j-1}}\frac{\delta {\cal S}}{\delta \phi} +\sum_{n=1}J_n^*
\frac{n(n+1)\cdots(n+j-1)-n(j-1)!}{(1+\phi)^{n+j}}.
}
\label{phi.1.3ppp}
\end{eqnarray}
For instance
\begin{eqnarray}&&
{\cal N}_2 = 
\Bigl[-J_0^*\frac{1}{(1+\phi)^2}
+\sum_{n=1}n(n+1)
J_n^*\frac{1}{(1+\phi)^{n+2}} \Bigr]\Bigr|_{J_0^*=-\frac{\delta \Gamma^{(0)}}{\delta\chi (x)}}
\nonumber\\&&
 {
 = \frac{1}{(1+\phi)}\frac{\delta {\cal S}}{\delta \phi} +\sum_{n=1}J_n^*
\frac{n^2}{(1+\phi)^{n+2}}
}
\label{hlc.-1}
\end{eqnarray}
and
\begin{eqnarray}&&
{\cal N}_3 = 
\Bigl[-J_0^*\frac{2}{(1+\phi)^3}
+\sum_{n=1}n(n+1)(n+2)
J_n^*\frac{1}{(1+\phi)^{n+3}} \Bigr]\Bigr|_{J_0^*=-\frac{\delta \Gamma^{(0)}}{\delta\chi (x)}}
\nonumber\\&&
 {
 = \frac{2}{(1+\phi)^{2}}\frac{\delta {\cal S}}{\delta \phi} +\sum_{n=1}J_n^*
\frac{n^2(n+3)}{(1+\phi)^{n+3}}.
}
\label{hlc.0}
\end{eqnarray}
Notice also the relations ($j>0$)
\begin{eqnarray}&&
\frac{\delta {\cal N}_j(x)}{\delta \phi(y)}
=\delta(x-y)\Bigl[- {\cal N}_{j+1}+ \frac{(j-1)!}{(1+\phi)^{j-1}}{\cal N}_2
\Bigr]
\nonumber\\&&
\frac{\delta {\cal N}_j(x)}{\delta \phi^*(y)}
=\frac{(j-1)!}{(1+\phi)^{j-1}}\frac{\delta ^2{\cal S}}
{\delta \phi^*(y)\delta \phi(x)}
.
\label{hlc.0.1}
\end{eqnarray}
Under rotations one has
\begin{eqnarray}&&
\rho[\alpha] {\cal N}_{j}=
i\alpha j! \frac{J_0^*\phi}{(1+\phi)^{j+1}}
+ i\alpha j! \frac{J_0^*+J_1^*}{(1+\phi)^{j+1}}
\nonumber\\&&
-i\alpha (j+1)!\frac{J_1^*\phi}{(1+\phi)^{j+2}}
-i\alpha\sum_{n=2}n \cdots(n+j)\frac{J_n^*\phi}{(1+\phi)^{n+j+1}}
\nonumber\\&&
+ i\alpha\sum_{n=2}n^2 (n+1)\cdots(n+j-1)\frac{J_n^*}{(1+\phi)^{n+j}}
- i\alpha\sum_{n=1}n\cdots(n+j) \frac{J_n^*}{(1+\phi)^{n+j+1}}
\nonumber\\&&
= 
ij!\alpha \frac{J_0^*}{(1+\phi)^{j}}
- i\alpha\sum_{n=1}n\cdots(n+j) \frac{J_n^*}{(1+\phi)^{n+j}}
\nonumber\\&&
+ i\alpha\sum_{n=1}n^2 (n+1)\cdots(n+j-1)\frac{J_n^*\phi}{(1+\phi)^{n+j}}
=- i\, j\, \alpha {\cal N}_{j}.
\label{al.5}
\end{eqnarray}
%

\subsection{Properties of the translations generator $\tau[\beta]$}
Consider now the properties under  translations.
Direct evaluation yields
\begin{eqnarray}&&
\tau[\beta]  \frac{\delta \Gamma^{(0)}}{\delta\phi (x)}
= \tau[\beta]  \Bigl[J^*
- \sum_{n=1}n \frac{J_n^*}{(1+\phi)^{n+1}}
\Bigr]
\nonumber\\&&
= -\beta \frac{\delta \Gamma^{(0)}}{\delta\phi (x)}\frac{1}{1+\phi}
+2 \beta\frac{J_1^*}{(1+\phi)^{3}}
-\beta \sum_{n=2}\Bigl[n (n-1)\frac{J_{n-1}^*}{(1+\phi)^{n+1}}
\nonumber\\&&
- n(n+1)\frac{J_n^*}{(1+\phi)^{n+2}}
\Bigr]
\nonumber\\&&
= -\beta \frac{\delta \Gamma^{(0)}}{\delta\phi (x)}\frac{1}{1+\phi}
\label{be.2}
\end{eqnarray}
or 
\begin{eqnarray}
\tau[\beta]  \frac{\delta \Gamma^{(0)}}{\delta\chi (x)}=0.
\label{be.3}
\end{eqnarray}
By using  
\begin{eqnarray}&&
\tau[\beta]  \frac{\delta {\cal S}}{\delta\phi^* (x)}
= \tau[\beta]  \Bigl[-(\Box+m^2)\phi +J+iF^\mu\partial_\mu\phi
+i\partial^\mu(F_\mu\phi) + K\phi
\Bigr]
\nonumber\\&&
=-(\Box+m^2)\beta  
- \int d^Dy\beta(y)\frac{\delta^2{\cal S}}{\delta \phi (y)\phi^* (x)}
+iF^\mu\partial_\mu\beta
+i\partial^\mu(F_\mu\beta)
\nonumber\\&&
 + K\beta =0,
\label{be.5}
\end{eqnarray}
one gets
\begin{eqnarray}&&
\tau[\beta]  \frac{\delta \Gamma^{(0)}}{\delta\chi^* (x)}
= \tau[\beta]  \frac{\delta {\cal S}}{\delta\chi^* (x)}
= \tau[\beta]  \Bigl[(1+\phi^*) \frac{\delta {\cal S}}{\delta\phi^* (x)}
\Bigr]
\nonumber\\&&
=(1+\phi^*)\tau[\beta]  \Bigl[ \frac{\delta {\cal S}}{\delta\phi^* (x)}
\Bigr]=0.
\label{be.4}
\end{eqnarray}
One gets also
\begin{eqnarray}
\tau[\beta]  \frac{\delta {\cal S}}{\delta\phi (x)} =0.
\label{be.5.1}
\end{eqnarray}
The transformation properties under translations of 
${\cal N}_0, {\cal N}_1,{\cal N}_j$ are
\begin{eqnarray}&&
\tau[\beta]  {\cal N}_0
=\tau[\beta]   J_0^*\ln(1+\phi)
+ \beta\frac{J_0^*}{(1+\phi)} - \beta \frac{J_0^*}{(1+\phi)}
\nonumber\\&&
- \beta\frac{J_1^*}{(1+\phi)^2}+\beta\sum_{n=1}n \frac{J_n^*}{(1+\phi)^{n+1}}
-\beta\sum_{n=2}n \frac{J_n^*}{(1+\phi)^{n+1}}
\nonumber\\&&
=-\tau[\beta]  \frac{\delta \Gamma^{(0)}}{\delta\chi(x)}\ln(1+\phi)
=0.
\label{be.6}
\end{eqnarray}
Similarly one shows that 
\begin{eqnarray}&&
\tau[\beta]  {\cal N}_1 =0
\label{be.7}
\end{eqnarray}
and
\begin{eqnarray}&&
\tau[\beta]  {\cal N}_j =0.
\label{be.8}
\end{eqnarray}
From the definitions (\ref{phi.1.3p}) and (\ref{phi.1.3ppp}) and 
with the use of the
identity (\ref{be.4}) it follows
\begin{eqnarray}&&
\tau[\beta]  {\cal N}^*_j =0, \quad j=0,\cdots.
\label{be.9}
\end{eqnarray}
%

\subsection{The algebra}
By using the previous results,  one can now prove the following
relation
\begin{eqnarray}
\Bigl[\rho[\alpha],\tau[\beta]\Bigr]
=-i \tau[\alpha\beta].
\label{calge.25p}
\end{eqnarray}
The most difficult terms are those involving $\frac{\delta }{\delta J_1^*}$\,.
Thus  I consider only those
\begin{eqnarray}&&
\Bigl[\rho[\alpha],\tau[\beta]\Bigr]
_{\frac{\delta }{\delta J_1^*}}
= \int d^D y \beta(y)\rho[\alpha] \frac{\delta \Gamma^{(0)}}{\delta\chi (y)}
\frac{\delta }{\delta J_1^*(y)}
\nonumber\\&&
-\int d^D x \,\,\,\alpha(x) \tau[\beta ]
[-i \frac{\delta \Gamma^{(0)}}{\delta\chi (x)} + i J_1^*]
\frac{\delta }{\delta J_1^*(x)}.
\label{alg.1}
\end{eqnarray}
Now  I use eqs. (\ref{lin.4.1p}), (\ref{al.3})  and (\ref{be.3}) and  I get
\begin{eqnarray}
\Bigl[\rho[\alpha],\tau[\beta]\Bigr]\Biggr|
_{\frac{\delta }{\delta J_1^*}}
= -i \int d^D x \,\,\alpha\beta \frac{\delta \Gamma^{(0)}}{\delta\chi (x)}
\frac{\delta }{\delta J_1^*(x)},
\label{alg.1.1}
\end{eqnarray}
which is the right term as it appears in eq. (\ref{bra.3}).

\section{One-loop counterterms }
\label{sec:olc}
The study performed in Section \ref{sec:wpc} shows that
at one-loop level the expected divergent ancestor amplitudes are
\begin{enumerate}
\item $K$-tadpole
\begin{eqnarray}
\frac{\delta\Gamma^{(1)}}{\delta K(x)} 
= \Delta_m(0)
\label{olc.1}
\end{eqnarray}
\item $J-J^*$ two-point function
\begin{eqnarray}
\frac{\delta\Gamma^{(1)}}{\delta J(x)\delta J^*(y)}
= \frac{i}{2} \Delta_m(x-y)^2
\label{olc.2}
\end{eqnarray}
\item $J-J_n^*$ two-point function
\begin{eqnarray}
\frac{\delta\Gamma^{(1)}}{\delta J(x)\delta J_n^*(y)}
= n^2\frac{i}{2} \Delta_m(x-y)^2
\label{olc.2.1}
\end{eqnarray}
\item $J_n-J_{n'}^*$ two-point function
\begin{eqnarray}
\frac{\delta\Gamma^{(1)}}{\delta J_n(x)\delta J_{n'}^*(y)}
= n^2{n'}^2\frac{i}{2} \Delta_m(x-y)^2
\label{olc.3}
\end{eqnarray}
\item $K-K$ two-point function
\begin{eqnarray}
\frac{\delta\Gamma^{(1)}}{\delta K(x)\delta K(y)}
= i \Delta_m(x-y)^2
\label{olc.4}
\end{eqnarray}
\item $F_\mu-F_\nu$ two-point function
\begin{eqnarray}
\frac{\delta\Gamma^{(1)}}{\delta F^\mu(x)\delta F^\nu(y)}
= i  \Delta_m(x-y)\su{\leftrightarrow}{\partial_\mu}
\su{\leftrightarrow}{\partial_\nu}\Delta_m(x-y).
\label{olc.5}
\end{eqnarray}

\end{enumerate}
By using the pole parts
\begin{eqnarray}&&
\Delta_m \simeq \frac{1}{(4\pi)^2} \frac{2}{D-4} m^2 
\nonumber\\&&
\Delta_m^2  \simeq i\frac{1}{(4\pi)^2} \frac{2}{D-4}
\label{olc.6}
\end{eqnarray}
and eq. (\ref{a.4}) in Appendix \ref{app:a},
one can write the counterterms at one loop
\begin{eqnarray}&&
\hat\Gamma^{(1)} = \Lambda^{D-4}\frac{1}{(4\pi)^2} \frac{2}{D-4}
\Bigl[m^2 {\cal I}_1
- \frac{1}{6}{\cal I}_2+ \frac{1}{2}{\cal I}_3+\frac{1}{2} {\cal I}_4
\Bigr]
\nonumber\\&&
\label{olc.7}
\end{eqnarray}
where
\begin{eqnarray}&&
{\cal I}_1= \int d^D x
(F^2-K)
\nonumber\\&&
{\cal I}_2= \int d^D x
(\partial_\mu F_\nu -\partial_\nu F_\mu )^2
\nonumber\\&&
{\cal I}_3= \int d^D x
(F^2-K)^2
\nonumber\\&&
{\cal I}_4= \int d^D x {\cal N}_2^* {\cal N}_2.
\label{olc.7.1}
\end{eqnarray}
Notice that the square modulus of $
{\cal N}_2$
describes the counterterms for all the two-point functions
$J-J^*$, $J-J_n^*$ and $J_n-J_{n'}^*$.
The rest of the divergent ancestor amplitudes
$F-F-K$ and $F-F-F-F$ are described by the local invariant
${\cal I}_3$. It is amazing that by expanding ${\cal I}_4$
in powers of the fields $\chi$ and $\chi^*$
one gets all the counterterms of the infinitely many descendant
amplitudes at one-loop.
\section{Some more local invariants}
\label{sec:local}
The results of the analysis of Section \ref{sec:sol}
on the properties of $\rho[\alpha] \,\,\, \tau[\beta]$ suggests 
many local invariants (i.e. solutions of
both $\rho[\alpha] {\cal X}=0$ and $\tau[\beta]  {\cal X}=0$).
Here is a partial list of them
\begin{eqnarray}&&
{\cal I}_5= \int d^D x
 \frac{\delta {\cal S}}{\delta\phi^* (x)}
 \frac{\delta {\cal S}}{\delta\phi (x)}
\nonumber\\&&
{\cal I}_6= \int d^D x
\frac{\delta \Gamma^{(0)}}{\delta\chi (x)}\frac{\delta \Gamma^{(0)}}{\delta\chi^*(x)}
\nonumber\\&&
{\cal I}_7= \int d^D x \Bigl[{\cal N}_0^* \frac{\delta\Gamma^{(0)}}{\delta\chi^*(x)}
+ 
{\cal N}_0
\frac{\delta \Gamma^{(0)}}{\delta\chi (x)}
\Bigr]
\nonumber\\&&
{\cal I}_8= \int d^D x \Bigl[\frac{\delta\Gamma^{(0)}}{\delta\chi^*(x)}
+ 
\frac{\delta \Gamma^{(0)}}{\delta\chi (x)}
\Bigr]
\nonumber\\&&
{\cal I}_{9}= \int d^D x
D^\mu \frac{\delta {\cal S}}{\delta\phi^* (x)} 
D_\mu^* 
\frac{\delta {\cal S}}{\delta\phi (x)}
\nonumber\\&&
{\cal I}_{10}= \int d^D x D^\mu{\cal N}_1^* D_\mu^*{\cal N}_1
\nonumber\\&&
{\cal I}_{11}= \int d^D x (\partial^\mu-2i F^\mu){\cal N}_2^* 
(\partial_\mu+2i F_\mu){\cal N}_2\,.
\label{al.4.x}
\end{eqnarray}
%
The invariants in eq. (\ref{al.4.x}) are excluded
at the one-loop level for various reasons: either they are
too high in dimensions (as $ {\cal I}_{9}, {\cal I}_{10}\, {\rm and} 
\,{\cal I}_{11}$)
or they contain source terms describing amplitudes that are not
divergent (as ${\cal I}_{8}$) or the  coefficients determined by the
divergent ancestor amplitudes are just zero 
(as $ {\cal I}_{5}, {\cal I}_{6} \,\,{\rm and} \,{\cal I}_{7}$). 
\par
To conclude the Section it is worth noticing that, 
according to textbook
renormalization, one is expected to  introduce in the
classical action all the local invariants 
$ {\cal I}_{1}, {\cal I}_{2}, \,\,{\cal I}_{3} \,\,{\rm and} \,{\cal I}_{4}$
with arbitrary parameters.
However, in doing so, the very starting point given by the classical
action (\ref{infinite.8}) is deeply changed. The perturbative expansion
is modified by a whole set of new vertices. The WPC is not anymore
valid, due to the presence of four derivatives interactions. 
%
\section{Higher-order-loop counterterms }
\label{sec:hlc}
It is outside the scope of the present paper to perform
higher loop calculations. However in this Section some features
of the two-loop amplitudes are briefly discussed in order to
have a glance on the whole strategy.
\par
The two-loop invariant local counterterms must obey the inhomogeneous
equations where the given term is (e.g. in eq. (\ref{infinite.14p.3}))
\begin{eqnarray}
-\Lambda^{-D+4}
\frac{\delta\hat\Gamma^{(1)}}{\delta \chi(x)}
\frac{\delta\hat\Gamma^{(1)}}{\delta J_1^*(x)}.
\label{hlc.1}
\end{eqnarray}
Only ${\cal I}_4$  contributes to the
inhomogeneous term in eq. (\ref{hlc.1}). The others in eq. (\ref{olc.7})
do not depend on $\chi$ and $J_1^*$. One gets
\begin{eqnarray}&&
-\Lambda^{-D+4}
\frac{\delta\hat\Gamma^{(1)}}{\delta \chi(x)}
\frac{\delta\hat\Gamma^{(1)}}{\delta J_1^*(x)}
\nonumber\\&&
= \frac{ \Lambda^{D-4}}{(4\pi)^4} \frac{1}{(D-4)^2}\frac{{\cal N}^*_2(x)}{(1+\phi(x))^2}
\frac{\delta}{\delta\phi (x)}
\Bigl[\int d^D y {\cal N}_2(y){\cal N}^*_2(y)
\Bigr] 
.
\label{hlc.5}
\end{eqnarray}
To find a solution of inhomogeneous equation one needs some
more algebra. It useful to define the operator
\begin{eqnarray}
{\cal D} \equiv \rho +i(\tau-\tau^*).
\label{hlc.6}
\end{eqnarray}
Eqs. (\ref{infinite.11.3}) and (\ref{infinite.14p.3}) 
imply that any counterterm  $\hat\Gamma^{(n)}$ must obey 
the homogeneous equation
\begin{eqnarray}
{\cal D} \hat\Gamma^{(n)} =0.
\label{hlc.7}
\end{eqnarray}
The following relations can be easily derived
\begin{eqnarray}&&
{\cal D}(y) {\cal N}^*_{2}(x) =i 2\delta(x-y) {\cal N}^*_{2}(x)
\nonumber\\&&
{\cal D}(y)(1+\phi(x))^k= i k \delta(x-y)(1+\phi(x))^k
.
\label{hlc.7}
\end{eqnarray}
Moreover by using
\begin{eqnarray}
[{\cal D}(y) , \frac{\delta}{\delta \phi(x)}]=
- i\delta(x-y) \frac{\delta}{\delta \phi(x)}
\label{hlc.8}
\end{eqnarray}
one can show that
\begin{eqnarray}
{\cal D}(y)\frac{\delta {\cal X}}{\delta \phi(x)}=
- i \delta(x-y)\frac{\delta{\cal X}}{\delta \phi(x)},\qquad
\forall {\cal X} \quad :~{\cal D}{\cal X}=0.
\label{hlc.9}
\end{eqnarray}
Equations (\ref{hlc.7}), (\ref{hlc.8}) and (\ref{hlc.9})
give 
\begin{eqnarray}&&
{\cal D}(x)\,\, \int d^D y \frac{{\cal N}^*_2(y)}{1+\phi(y)}
\frac{\delta}{\delta\phi (y)}
\Bigl[\int d^D z {\cal N}_2(z){\cal N}^*_2(z)
\Bigr] 
=0
\nonumber\\&&
{\cal D}(x)\,\, \int d^D y \frac{{\cal N}^*_2(y)}{(1+\phi(y))^2}
 {\cal N}_2(y){\cal N}^*_2(y)
=0.
\label{hlc.10}
\end{eqnarray}
With a further identity
\begin{eqnarray}
[\tau(x),\frac{\delta}{\delta\phi (y)}]
=- \delta(x-y)(1+\phi(x)){\cal N}_2(x)
\frac{\delta}{\delta J_1^* (x)},
\label{hlc.10.1}
\end{eqnarray}
it  is straightforward to verify that
\begin{eqnarray}&&
\tau(x) \,\,\int d^D y \Biggl\{\frac{{\cal N}^*_2(y)}{1+\phi(y)}
\frac{\delta}{\delta\phi (y)}
\Bigl[\int d^D z {\cal N}_2(z){\cal N}^*_2(z)
\Bigr]
- \frac{{\cal N}^*_2(y){\cal N}_2(y)}{2(1+\phi(y))^2}{\cal N}^*_2(y)\Biggr\}
\nonumber\\&&
 =
- \frac{{\cal N}^*_2(x)}{(1+\phi(x))^2}
\frac{\delta}{\delta\phi (x)}
\Bigl[\int d^D z {\cal N}_2(z){\cal N}^*_2(z)
\Bigr]
\nonumber\\&&
- \frac{{\cal N}^*_2(x)}{(1+\phi(x))^3}{\cal N}_2(x){\cal N}^*_2(x)
+\frac{{\cal N}^*_2(x)}{(1+\phi(x))^3}{\cal N}_2(x){\cal N}^*_2(x)
\nonumber\\&&
=- \frac{{\cal N}^*_2(x)}{(1+\phi(x))^2}
\frac{\delta}{\delta\phi (x)}
\Bigl[\int d^D z {\cal N}_2(z){\cal N}^*_2(z)
\Bigr],
\label{hlc.11}
\end{eqnarray}
i.e. a solution of eq. (\ref{infinite.14p.3}) has been found. 
Eq. (\ref{hlc.10})
guarantees that also eq. (\ref{infinite.11.3}) is
satisfied.
\par
To complete the two-loop analysis one needs to evaluate
the divergent parts of
a (finite) set  of  ancestor amplitudes, enough
 to fix the coefficient of the solution in eq. (\ref{hlc.11}) and 
the homogeneous part of $\hat\Gamma^{(2)}$.

\section{Conclusions}
\label{sec:concl}
The use of the LFE's, derived from the invariance
properties of the path integral measure, of the hierarchy and of
the WPC allows a complete classification of the divergent
amplitudes for the scalar complex free field theory
in polar coordinates. Minimal subtraction in dimensional
regularization, enforced by the prescription of pure pole
subtraction, agrees with this structure  and thus
provides a perfect procedure for the symmetric subtraction
(i.e. preserving the LFE's)
of all the infinities. Thus the theory can be made finite at $D=4$.
\par
In this paper the LFE's are derived, the hierarchy
is proven and discussed, the WPC is illustrated as an
important tool for the analysis of the divergent ancestor
amplitudes. Finally the local solutions of the LFE's 
at one loop are discussed in some details and the counterterms
$\hat\Gamma^{(1)}$ are established. It is argued that
conventional renormalization procedure is not a viable method
for the use of polar coordinates, while the approach
presented in this paper yields a perturbative expansion
which is finite and consistent. It is briefly illustrated
(at the one-loop-level)
that the original field in cartesian coordinates ($\phi$) has the
correct two-point-function. Moreover a solution of the
inhomogeneous LFE's is derived for the two-loop counterterms.
One can conclude with good confidence that the existence
of the coordinates transformation for $D=4$
is proven by construction.

\section*{Acknowledgments}
I am honored to thank the warm hospitality of the
Center for Theoretical Physics at MIT, Massachusetts, where I 
had the possibility to work  on the present paper.
I thank Andrea Quadri for stimulating discussions.

\appendix

\section{$F_\mu-F_\nu$ two-point-function}
\label{app:a}
The one loop contribution to the two-point function $F_\mu-F_\nu$ is
as in scalar QED
\begin{eqnarray}
&& 
\Gamma^{(1)}_{F^{\mu}F^{\nu}}(p) =-i \int \frac{d^D k}{(2\pi)^D} 
\left [
\frac{(2k+p)_{\mu}(2k+p)_{\nu}}{(k^2-m^2)[(p+k)^2-m^2]} 
- 2\frac{g_{\mu\nu}}{k^2-m^2}
\right]
\nonumber\\&&
-i \int \frac{d^D k}{(2\pi)^D} 2\frac{g_{\mu\nu}}{k^2-m^2}
\nonumber\\&&
= -i 4(p_\mu p_\nu-p^2g_{\mu\nu}) \int \frac{d^D k}{(2\pi)^D} 
\int_{-\frac{1}{2}}^{\frac{1}{2}}dy 
\frac{ y^2}{[k^2-C]^2} 
-  2 g_{\mu\nu}\Delta_m(0)
\label{a.1}
\end{eqnarray}
where 
\begin{eqnarray}&&
C\equiv m^2 - p^2(\frac{1}{4}-y^2)
\nonumber\\&&
\Delta_m(0) = i\int \frac{d^D k}{(2\pi)^D}\frac{1}{k^2-m^2}
\label{a.2.0}
\end{eqnarray}
and
use has been made of the identity
\begin{eqnarray}
\int d^D k \frac{k^2}{[k^2-C]^2}
=\frac{D}{D-2}  \int d^D k   \frac{C}{[k^2-C]^2}.
\label{a.2}
\end{eqnarray}
%
The integration on the momentum gives
\begin{eqnarray}
&& 
\Gamma^{(1)}_{F^{\mu}F^{\nu}}(p) =
4(p_\mu p_\nu-p^2g_{\mu\nu}) \int_{-\frac{1}{2}}^{\frac{1}{2}}dy y^2
\frac{1}{(4\pi)^{\frac{D}{2}}} 
 \frac{\Gamma(2-\frac{D}{2})}
{\Gamma(2)}[C]^{\frac{D}{2}-2}
\nonumber\\&&
-  2 g_{\mu\nu}\frac{1}{(4\pi)^{\frac{D}{2}}} 
 \frac{\Gamma(2-\frac{D}{2})}
{(1-\frac{D}{2})\Gamma(1)}[m^2]^{\frac{D}{2}-1}
\label{a.3}
\end{eqnarray}
The pole part is then
\begin{eqnarray}
&& \Lambda^{-D+4}
\Gamma^{(1)}_{F^{\mu}F^{\nu}}(p)\Bigr|_{\rm pole~part} =
\frac{2}{3} \frac{1}{(4\pi)^2} \frac{1}{D-4}(p^2g_{\mu\nu}-p_\mu p_\nu)
\nonumber\\&&
-  4\frac{1}{(4\pi)^2} \frac{1}{D-4}
 m^2 g_{\mu\nu}.
\label{a.4}
\end{eqnarray}
%

\bibliography{reference}

\end{document}